\documentclass[a4paper, 11pt]{article} 

\usepackage{graphicx} 
\usepackage{wrapfig} 

\usepackage{natbib}

\makeatletter
\renewcommand\@biblabel[1]{\textbf{#1.}} 
\renewcommand{\@listI}{\itemsep=0pt} 

\renewcommand{\maketitle}{ 
\begin{flushright} 
{\LARGE\@title} 

\vspace{50pt} 

{\large\@author} 
\\\@date 

\vspace{40pt} 
\end{flushright}
}

\title{\textbf{Decentralization in Digital Societies}\\ 
A Design Paradox\footnote{This essay is based on material presented at the 2016 Salon Festival, Maloja Palace, Switzerland: \emph{In Pursuit of the Beautiful Soul, The Public Sphere Salons}, https://www.publicspheresalons.com}} 

\author{\textsc{Evangelos Pournaras} 
\\{\textit{School of Computing, University of Leeds, Leeds LS2 9JT, UK}}} 

\begin{document}

\maketitle 



\begin{abstract}
Digital societies come with a design paradox: On the one hand, technologies, such as Internet of Things, pervasive and ubiquitous systems, allow a distributed local intelligence in interconnected devices of our everyday life such as smart phones, smart thermostats, self-driving cars, etc. On the other hand, Big Data collection and storage is managed in a highly centralized fashion, resulting in privacy-intrusion, surveillance actions, discriminatory and segregation social phenomena. What is the difference between a distributed and a decentralized system design? How "decentralized" is the processing of our data nowadays? Does centralized design undermine autonomy? Can the level of decentralization in the implemented technologies influence ethical and social dimensions, such as social justice? Can decentralization convey sustainability? Are there parallelisms between the decentralization of digital technology and the decentralization of urban development?
\end{abstract}

\hspace*{3,6mm}


\vspace{30pt} 


\section*{Rhizome of the Big, Suppression of the Small}

Are data actually "Big" in digital societies? Scratching the surface of Big Data is used as a philosophical narrative for an in-depth comprehension of the buzzword, the actual design it conveys and the techno-socio-economic implications of this design. 

Information and Communication Technologies (ICT) such as Internet of Things, ubiquitous and pervasive computing, wearable devices and other have brought paramount opportunities for sustainable digital societies in application domains such as Smart Cities, Smart Grids and ambient-assisted living. Digital societies provide functionality and services that reason based on empirical data. The vast majority of these data can be generated \emph{locally} by each citizen who uses the aforementioned ICT technologies. Given that nowadays most citizens in developed and developing countries have access to some of these technologies, the data generation is highly \emph{participatory} and \emph{decentralized by design}. The data corresponding to each citizen are only a small fraction of the total data generated at a global scale. Therefore, the proportion of data corresponding to each citizen is nowadays magnitudes lower compared to the past when the participatory actions based on ICT were minimal and only large corporations could have access to these costly technologies. We ultimately live in an era of "Small Data".

So what makes the "Small Data" "Big"? Does Big Data convey a misconception or a paradox? Big Data is actually a rhizome of massive data collection practices governed by large corporations or governments whose systems design is highly detached from the decentralized nature of data generation. This practice suppresses and eventually undermines the inherent decentralized design of digital societies. Although Big Data technologies claim decentralized/distributed processing of data using programming models such as MapReduce, these technologies are actually deployed and used in highly centralized settings. Data are collected, stored and processed in large energy-intensive data centers, over which citizens have no control and authority. Distributed data processing within this highly centralized setting exclusively serves corporate performance and competitiveness. However, given the current economic arena, only a few powerful business players can invest on such expensive computational resources. This results in a cascade of centralization and power concentration as a tactical utility\footnote{\citet{Cummings1995} recalls former organization theorists with this view for the future digital societies.} mingled in technical, social, business, economic and political realities. The sustainability and cohesion of digital societies comes in question.

\section*{The Ongoing Battle Behind the New Manifestation}

The debate on centralized vs. decentralized design dates back to non-digital societies and its existence has philosophical relevance and significance. \citet{Cummings1995} relies on semantic decomposition to argue that the two terms are a binary undecidable opposition. They cannot be conceptualized apart from each other due to the intrinsically divided logic of writing. This creates inherently cyclic dynamics in the perceptions between centralization and decentralization. This philosophical view has reflections in empirical observations on fiscal, administrative, regulatory, market and financial centralization/decentralization of public services~\citep{Vries2000,Ahmad2005}. It is even pointed out that the same arguments are used to support either centralization or decentralization and that opposing arguments appear to support the same view among different countries. These contradicting views also have ideological origins, for instance, references to decentralization swing over anarchism, libertarian socialism and even neo-liberalism. 

\citet{Gershenson2005} illustrate the perspective of complexity science that moves beyond distinction conservation of classical sciences~\citep{Heylighen1989} and introduces the indeterminacy in which observations or distinctions made by observers in different contexts can vary. Beyond the prevalent conceptual applicability of indeterminacy in quantum mechanics, the indeterminancy between centralization and decentralization becomes more apparent when studying topological and spectral properties of complex networks representing techno-socio-economic systems~\citep{Provan2008,Strogatz2001,Boccaletti2006,Albert2002}.

\section*{Cascade Effects of Design}

Significant challenges that digital societies face nowadays stem from their design. For example, practices of privacy violation are a major concern in the Big Data era. Privacy can be violated (i) as a result of low citizens' awereness about the implications of giving away their personal data or (ii) by advanced inference techniques applied to partial/incomplete citizens' data. In both cases, centralization plays a key role. These privacy violations are a structural effect originated from the centralized design in information management. 

In the former case, complex privacy settings and policies in data collection are a mainstream that keep citizens under-informed about which of their personal data are collected and how they are used. Even when some privacy control is given back to citizens, this is counter-intuitively institutionalized and determined by the centralized authority that collects the data, the same potential violator of privacy. The notion of conflict of interest does not apply in this case. This centrally determined privacy control can ironically turn out be deceiving or opportunistic as choices about privacy are personal data collected as well. For example, the control of which friends can see a picture uploaded in a centralized social network reveals a level of trust, a ranking of human relationships camouflaged under a notion of privacy determination. At the end, most social networks may allow each individual to choose what is shared with everyone else except themselves. In conclusion, unless citizens self-institute and self-determine information sharing, centralized data collection cannot by design contribute to citizens' awareness in privacy and can even further violate their privacy. 

In the latter case of privacy intrusion via inference, it is again the centralized design that opens up ways to violate privacy. Inference is usually performed by deducing some missing or new type of information by using analysis of data sources. For example, identifying the TV channel and audio-visual content does not require the explicit reveal of this information by household residents. Surprisingly, it can be also inferred with high accuracy using household energy consumption data captured by smart meters~\citep{Greveler2012}. Privacy threats by inference are even more challenging for citizens to perceive, and therefore, to be aware of. Usually, privacy policies do not explicitly reflect on such threats. It is when different collected data streams are centralized and processed by powerful computational resources that unlimited inference opportunities arise. When data remain distributed and under citizens' control, inference is either literally or computationally infeasible. Decentralization entails a significant level of privacy-by-design, and can be adopted as a tactical utility for privacy-preservation. 

Privacy intrusion has a cascade of implications on autonomy of decision making, individuals' freedom and therefore, democracy~\citep{Helbing2015}. In a digital society of centralized information systems, new powerful ways of surveillance, discrimination, manipulation of public opinion and totalitarian e-governance emerge. Highly commercialized recommender systems or over/under-regulated computational markets often lack of a legitimate transparent access to citizens' data. As a result, the semiotics of information in opinion formation and decision-making are fundamentally altered~\citep{Eco2014}.

\section*{The Oxymoron of Sustainability}

Centralization also has an environmental impact. For example, the carbom emmisions of datacenters account for 14\% of the ICT footprint~\citep{Webb2008}, 2\% of all electricity usage in the USA and 1.3\% globally~\citep{Brown2008}. There is an active ongoing research on energy efficiency and savings of centralized computing infrastructures~\citep{Beloglazov2011}, however, the energy consumption of data centers continues to grow~\citep{Brown2008}. 

Energy efficiency in data centers cannot justify sustainability as the underlying environmental manifestation of the centralized design smolders unnoticed. If privacy could be preserved, data centers might not be needed at first place, or at least to the scale they are required nowadays. Beyond the ethical dimension, privacy violations such as the ones illustrated earlier have a measurable environmental impact as they require storage and processing capacity. Even if these computational resources are environmental-friendly, sustainability remains an oxymoron. Moreover, the need for a large-scale use of centralized data centers can be further limited if the underutilized disk space and processing capacity of personal computers and other distributed computational resources are explored~\citep{Benet2014,Swan2015}. Decentralizing the energy efficiency by focusing on environmental-friendly end-user technology can be a more effective and sustainable approach~\citep{Wang2009,Nurminen2008,Pantazis2013,Pournaras2013b,Pournaras2014}. 

The design bond between physical and digital finds another manifestation in the development of rural and urban environments. The centralization of information systems results in large ICT corporations physically close to administrative centers of cities, where they can sustain their business activities. This results in a further alienation of rural areas and losses of their competitive advantages. Undoubtedly and regardless of the design of information systems, citizens can benefit from higher quality of public services supported by digital means~\citep{Kostakis2015}. However, rather than Smart Towns or Smart Villages, it is no wonder that Smart Cities are the mainstream nowadays. Although the status quo suggests the city as the incubator of innovation, a more physiocratic view would mandate the repatriation of the innovation outcome in rural areas for reflecting the benefits to real economy and growth~\citep{Heinonen2013}. Such considerations are highly applicable in countries of the European South affected by the economic crisis and especially Greece that has a high level of urbanization, nevertheless an economy relying on primary sector of the economy.

\section*{Claiming the `Self'}

\citet{Eco2014} argues that true control in communication comes from the actual control of information meaning and its interpretation. This turns information from an instrument for producing economic merchandise into a chief merchandise. The tactical centralization in the Big Data era creates unlimited opportunities for control over meaning and its interpretation. The suppression of the inherent decentralized design of digital societies, along with the magma of power concentration by the centralization of information systems undermines the 'self' of self-instituting societies. Consequently, the foundations of democracy are undermined, as Castoriadis sees to the self-instituting societies the dawn of democracy back to ancient Greece~\citep{Castoriadis1983,Castoriadis1991}. 

This discussion does not imply that decentralization is a panacea and centralized design the cause of an upcoming dystopian future. Decentralized systems such as peer-to-peer networks have been criticized for the security holes, free-riding or illegal content sharing~\citep{Wallach2003}. Several of these issues are addressed by new novel decentralized technologies such as blockchain~\citep{Swan2015}, while others are a result of the existing well-established economic and political interests opposing a transition towards decentralization. Distinguishing between a weak outcome because of the transition to decentralization and a weak outcome because of a fundamental flaw in the actual decentralized design is a challenge to be addressed~\citep{Ahmad2005}. 

There is a plethora of applications in which decentralized information systems are an alternative or a natural fit within the domain applied. For example, decentralized micro-generation of energy empowers citizens to be both consumers and producers. Centralized computations for matching energy supply and demand in this dynamic decentralized environment can undermine privacy and autonomy as discussed earlier. In contrast, the reliability of Smart Grids can improve via self-organizing multi-agent systems running decentralized optimization mechanisms. Decentralization does not only contribute to cost-effectiveness but also to a welfare by minimizing human discomfort and maximizing social fairness~\citep{Pournaras2014,Pournaras2014b}. Similarly, data analytics are not a monopoly of Big Data systems. Measurements can also be performed in a fully decentralized fashion as a public good using collective intelligence distributed over computational resources of participatory citizens~\citep{Pournaras2018,Pournaras2015,Jesus2015}. 

Although the battle of decentralization in the Big Data era may resemble a digital guerrilla warfare, this battle is actually the claim of the missing `self' from self-instituting digital societies, the claim of a digital democracy worth pursuing.

\bibliographystyle{abbrvnat}

\bibliography{decentralization}

%
%
%
%
%
%
%
%


\end{document}